# Indian Derivatives Market Evolution and Challenges

***Shruthi B.C**
****Dr. N. Suresh**

*Research Scholar, Jain University, Bangalore.
**Professor & Director of Dayananda Sagar College of Management Studies

## Abstract

The study is conducted to establish the framework for comparting the relation in performance of the derivatives of BSE and NSE in India and to analyse the relationship of derivatives with cash market and the market volatility. The exchange traded equity derivatives were considered for the study. It was found that the performance of derivatives in NSE is lot higher than BSE and the NSE is on par with the global exchanges compared to BSE in terms of in terms of the number of contracts traded for Stock Index Options and Futures and also Stock Futures. Hence derivatives market need to strengthen further with the number of contracts traded and turnover in all the derivative instruments with more strong regulations and robust framework protecting the interest of the investors. This study enables Derivative Industry to progress towards its goals and objectives in a more efficient way.

## Introduction

Following the growing instability in the financial markets, the financial derivatives gained prominence after 1970. In recent years, the market for financial derivatives has grown in terms of the variety of instruments available, as well as their complexity and turnover. Financial derivatives have changed the world of finance through the creation of innovative ways to comprehend, measure, and manage risks. India's tryst with derivatives began in 2000 when both the NSE and the BSE commenced trading in equity derivatives. India's experience with the equity derivatives market has been extremely positive. India is one of the most successful developing countries in terms of a vibrant market for exchange-traded derivatives. This reiterates the strengths of the modern development in India's securities markets, which are based on nationwide market access, anonymous electronic trading, and a predominant retail market. There is an increasing sense that the equity derivatives market plays a major role in shaping price discovery. The present paper is descriptive in nature and based on the secondary data encompasses the prominence of derivative market and many issues underlying which need to be immediately resolved to enhance the investors' confidence in the Indian derivative market.

## Hypothesis:

1. There is a relation in the performance of Index and Stock Futures and Options of NSE and BSE.
2. Derivatives turnover influences the turnover in the Cash and Volatility segment of Indian Market.
3. Indian derivatives market growth trend is in line with the global exchanges.

## Scope of the study:

In India, the emergence and growth of derivatives market is relatively a recent phenomenon. Since its inception in June 2000, derivatives market has exhibited exponential growth both in terms of volume and number of contract traded. The market turnover of derivatives of NSE and BSE has grown from Rs.4,038 Cr. in 2000-2001 to Rs.3, 21,58,208 Cr. in 2012-13. Within a short span of twelve years, derivatives trading in India has surpassed cash segment in terms of turnover and number of traded contracts. Hence it is important to critically view the Issues and the challenges revolving around the derivative industry and the Regulations governing them.

**Plan of analysis**

The statistical data listing of all the variables collected are in tables, wherein mean and standard deviation as been used for continuous variables such as Turnover, Number of Contracts, etc. The pair t-test was applied to test the difference of means in two variables, and one-way ANOVA was applied for more than two means. All statistical tests are two-tailed, at $\alpha=0.05$. All trends were analysed and studied by linear regression equation. Statistical Analysis was done through MS-EXCEL (Advance), as the data-base was prepared in MS-EXCEL. Data validation was done by using data validation tool of MS-EXCEL.

**Data Analysis**

Graph showing the yearly comparison of the Cash & Derivatives Market Turnover of BSE &NSE

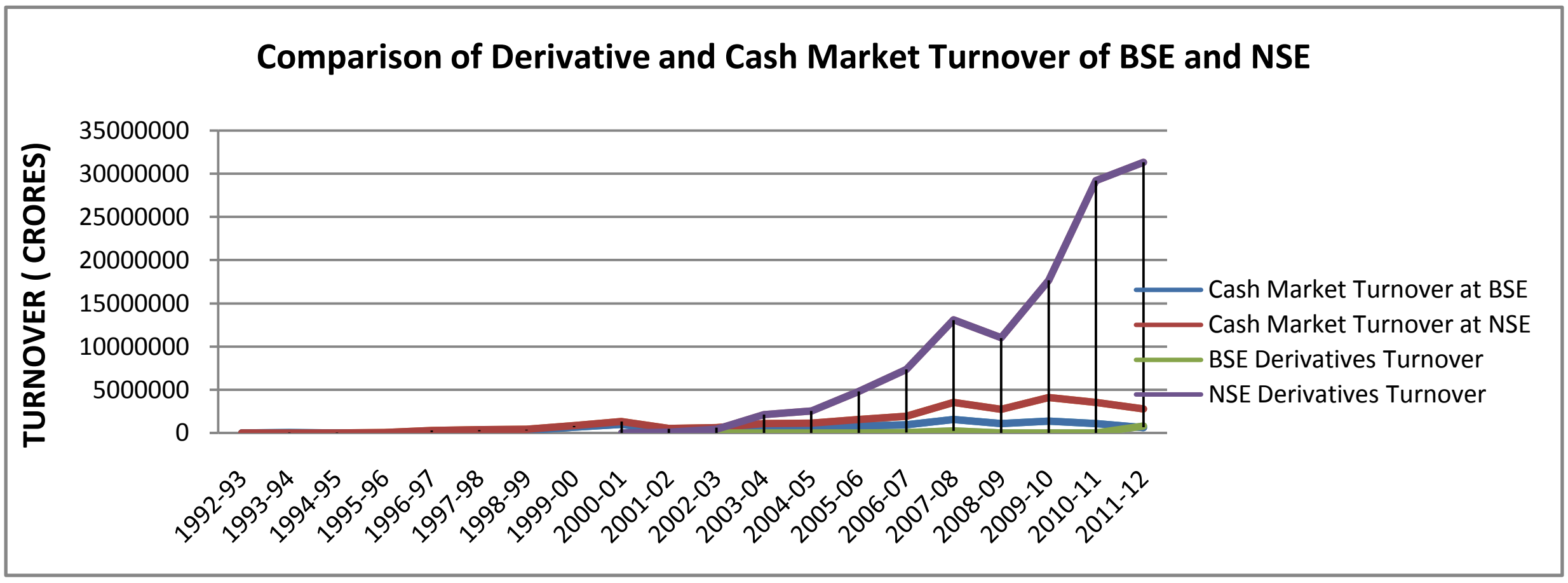

Source: (sebi.gov.in) Graph: 1

The above graph shows the cash market and derivatives turnover from 1992 to 2012. BSE cash market includes BSE Sensex commenced from January 2, 1986 and BSE-100 Index which commenced from April 3, 1984 and NSE cash market includes S&P CNX Nifty Index which commenced from November 3, 1995 and CNX Nifty Junior which commenced from November 4, 1996. The cash market of BSE was Rs.45,696 crores in 1992-93 and increased till Rs. 10,00,032 crores in 2000-01 the year derivatives was introduced. Derivatives turnover stood at Rs. 1673 crores in 2000-01 and Rs. 808476 crores where it surpassed the turnover of Cash Market. The NSE derivative market by the figure itself reflects the high margin increase in from 2002-03 where the derivatives turnover was Rs. 439866 crores and cash market turnover was Rs.314073 crores.

Graph showing the daily comparison of the Cash segment and Derivatives Turnover of BSE and NSE

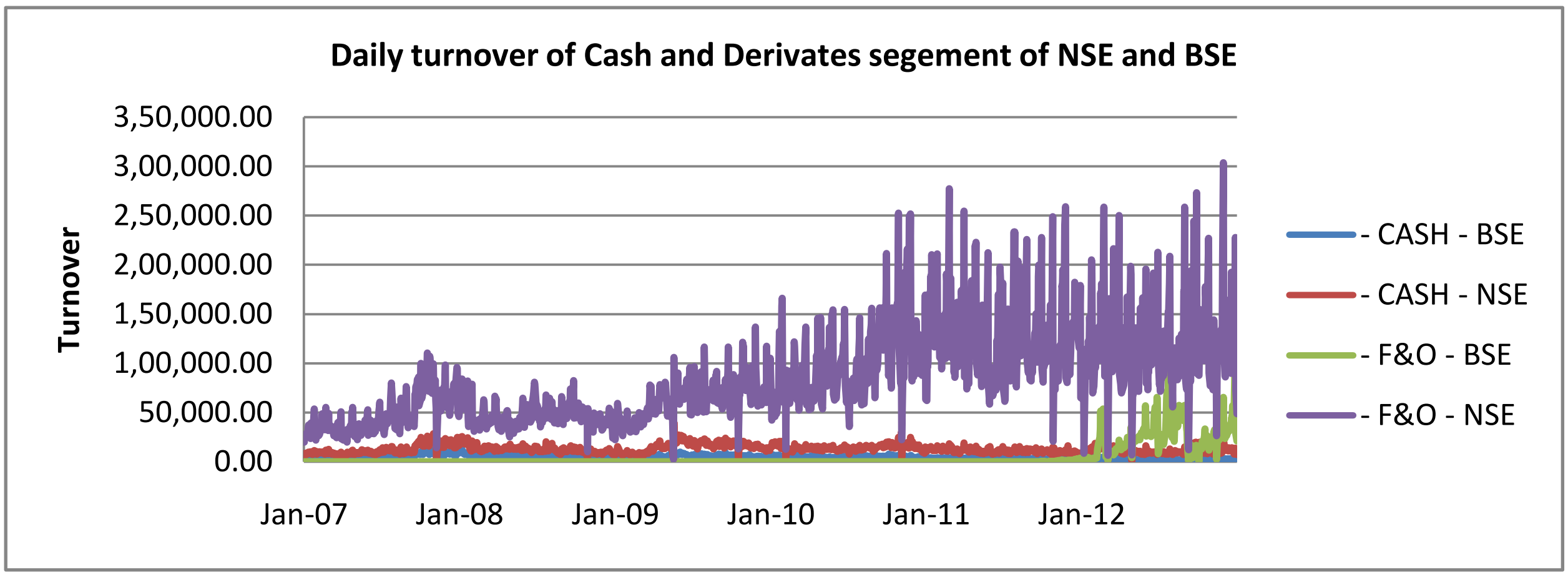

Source: moneycontrol.com Graph: 2

The above graph refects the daily trend growth of Cash and F&O segment turnover where the cash market turnover as on 2nd Jan 2007 was Rs.3, 380.94 crores but the Futures and Options turnover was gained only from 24th Jan 2011 which was Rs. 0.03 crores but it surpassed the cash market by 31st December 2012 where the turnover of derivatives was Rs. 20,944 crores but the cash market turnover was Rs. 1,781.61 crores. The F&O segment of NSE has always surpassed the Cash market with the date beginning from 2nd Jan 2007 with Rs. 19,957.26 crores to Rs. 48,894.61 crores on 31st December 2012 and the Cash market turnover beginning with Rs. 5,938.27 crores to Rs. 7,547.27 as on 31st December 2012.

Graph showing Monthly comparison of the turnover of Index Futures in BSE and NSE

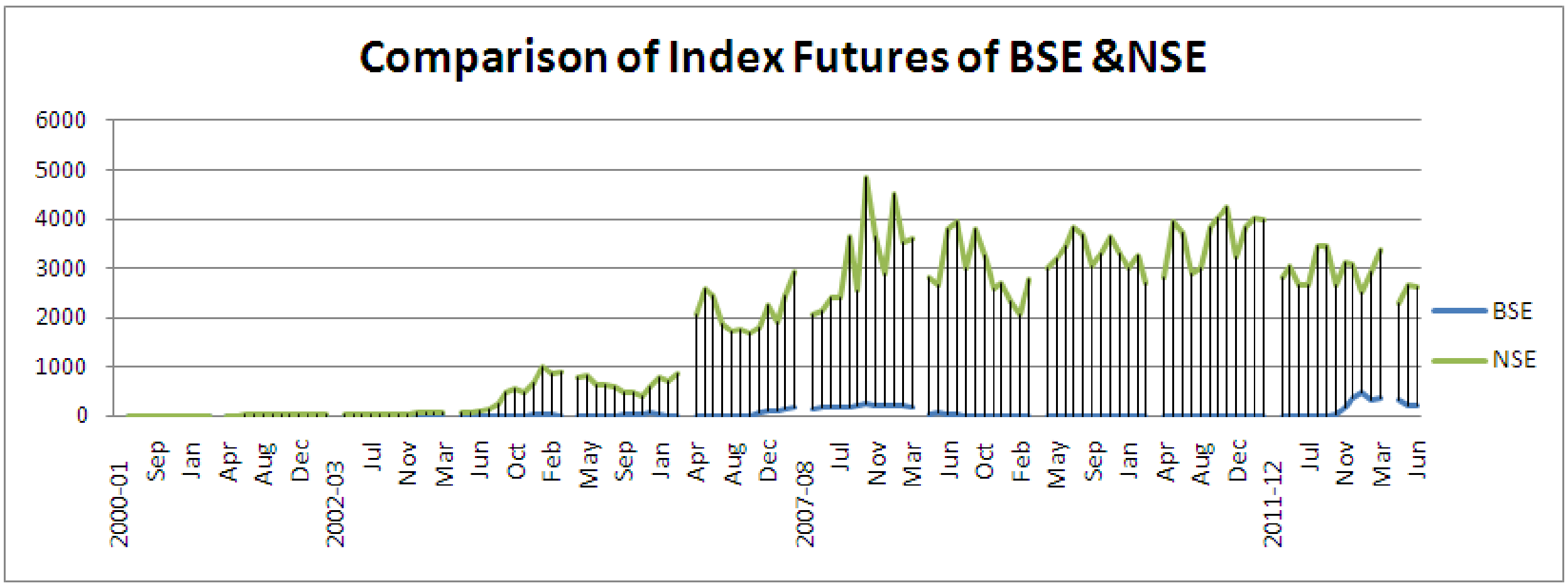

Source: SEBI Graph: 3

Handbook of Statistics on Indian Securities Market 2012

It is evident from the above graph that there is huge gap in the turnover between the two Stock Exchanges when compared on Index Futures.

Graph showing Monthly comparison of the turnover of Index Options in BSE and NSE

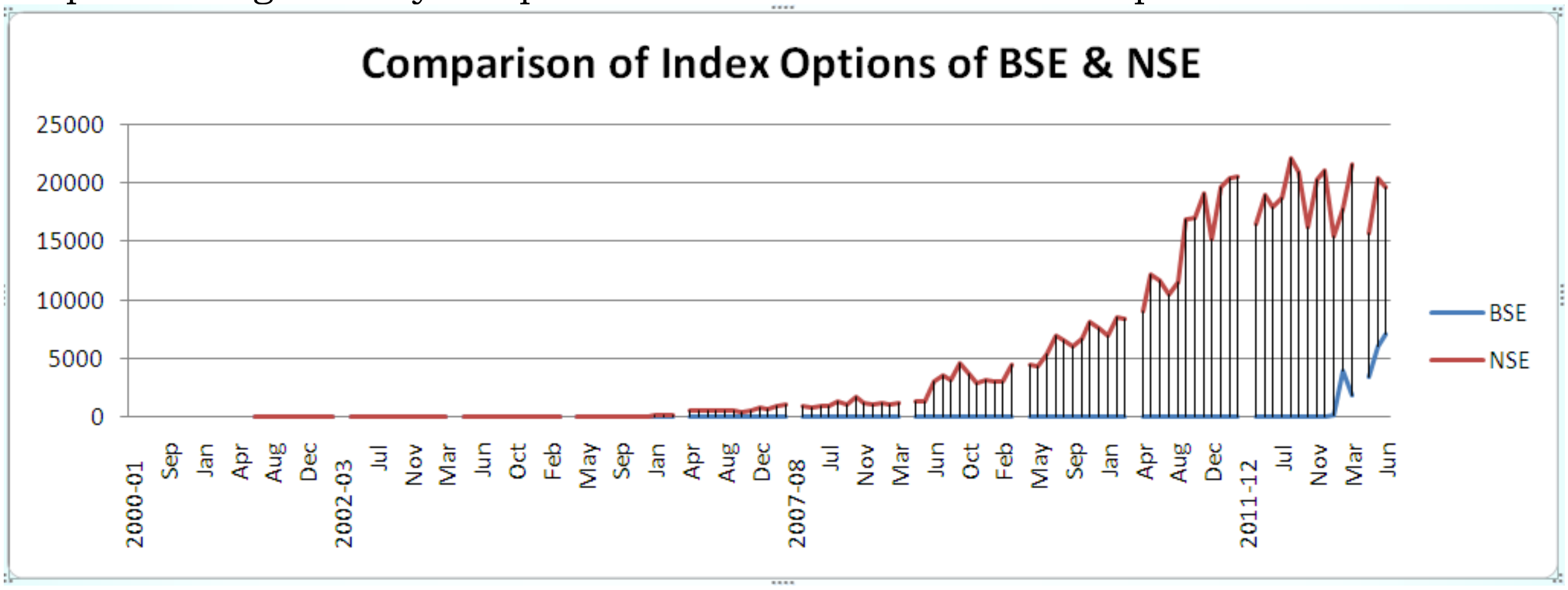

Source: SEBI Graph: 4

Handbook of Statistics on Indian Securities Market 2012

It is evident from the above graph that there is huge gap between the turnovers of the two Stock Exchanges when compared on Index Options too.

Graph showing Monthly comparison of the turnover of Stock Options in BSE and NSE

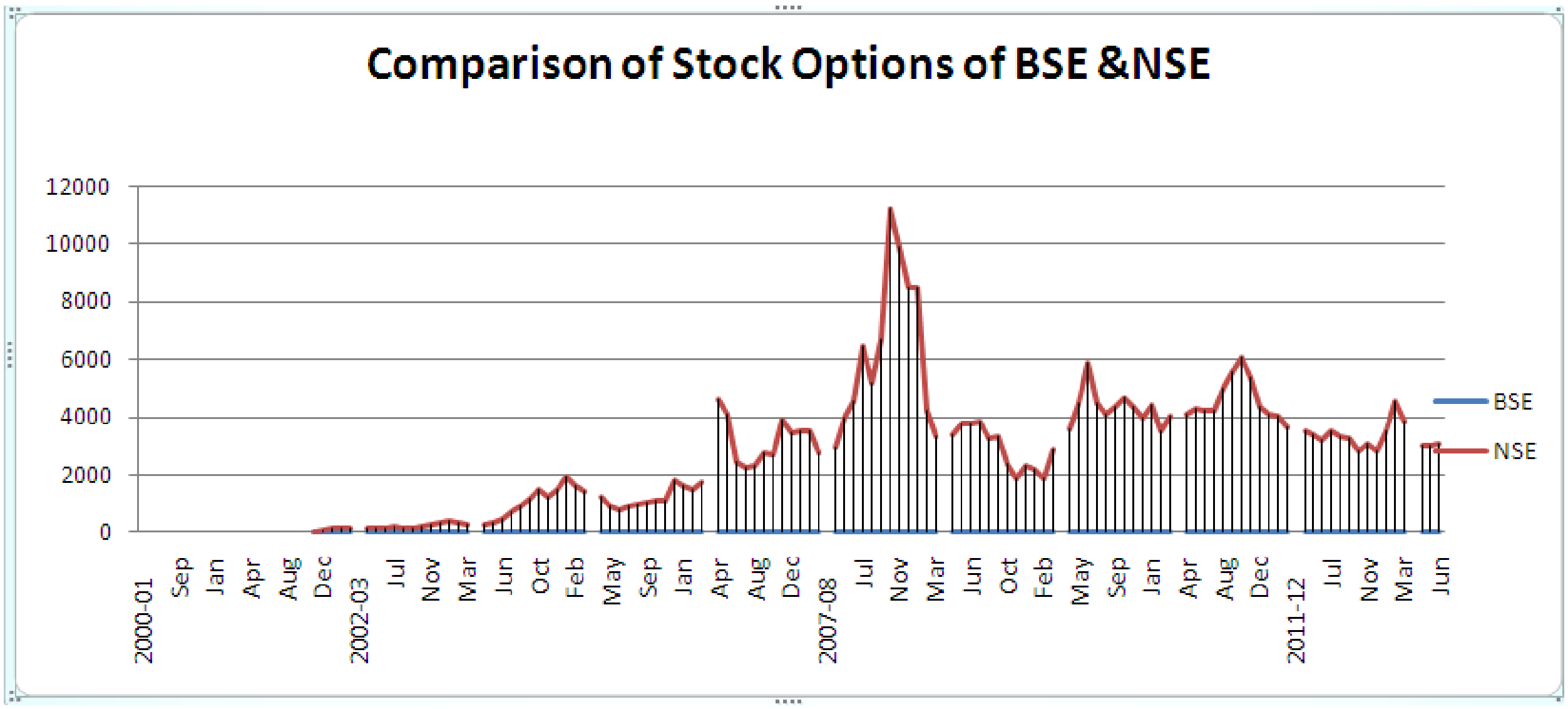

Source: SEBI Graph: 5

Handbook of Statistics on Indian Securities Market 2012

The above graph also suggests there is huge difference in the monthly turnover of Stock Options in BSE and NSE.

Graph showing Monthly comparison of the turnover of Stock Futures in BSE and NSE

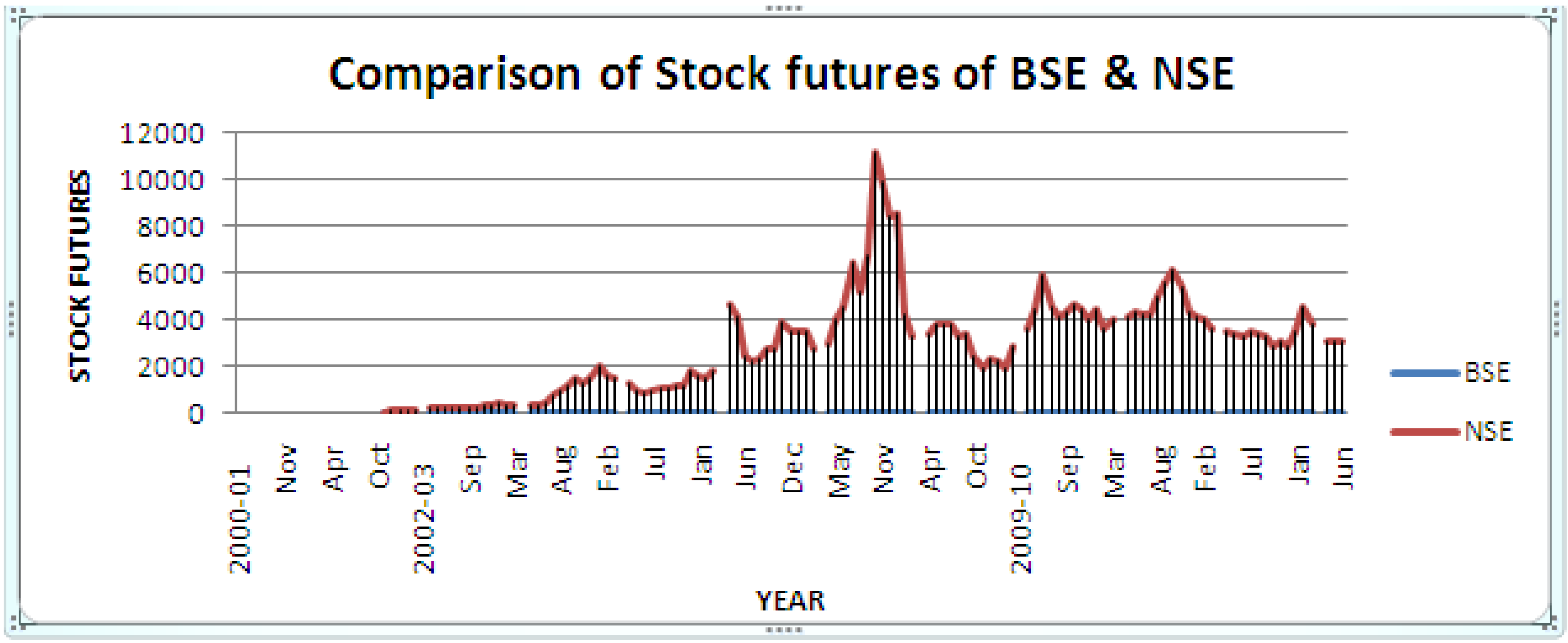

Source: SEBI Graph: 6

The above suggests the same like previous graph that there is huge difference in the monthly turnover of Stock Futures in BSE and NSE.

Graph showing the Volatility of Major Indices of Indian Captial Market

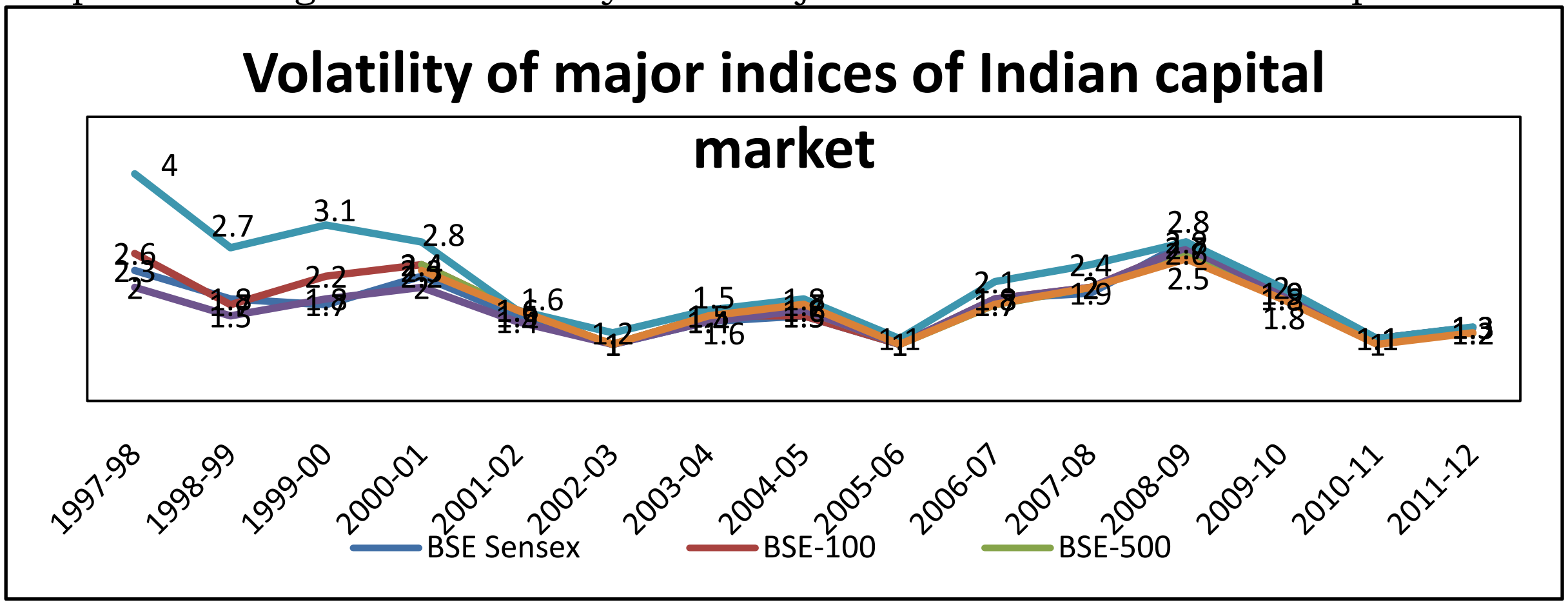

Source: (sebi.gov.in) Graph: 7

The price volatility of all the indices of BSE and NSE move in the similar trend over the years but the volatility shows a decrease in 2000-01 which could be the reason of introduction of derivatives and also may be may be due to many other factors, including better information dissemination and more transparency and increase in 2008-09 which could be due to the Global financial crisis.

Graph showing the comparison of Index Futures of the Global exchanges with BSE and NSE

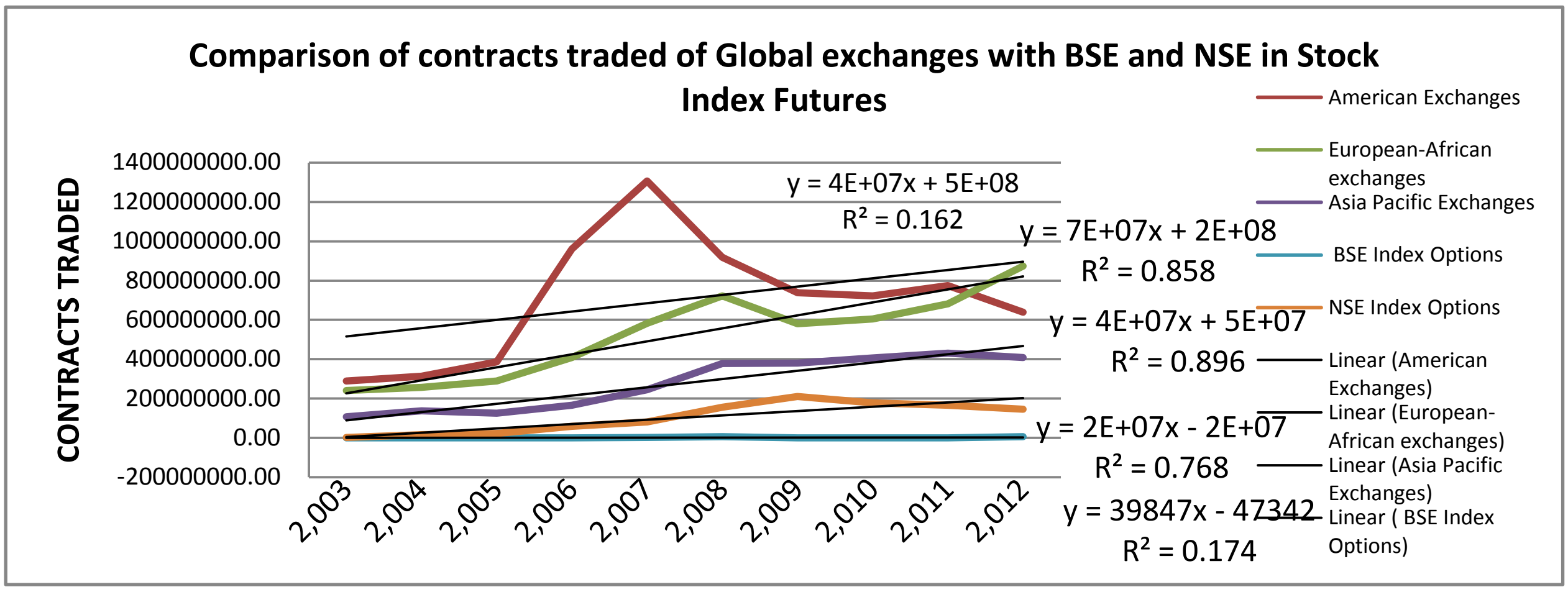

Source: World Federation of Exchanges Graph No: 8

Graph showing the comparison of Stock Index Options of the Global exchanges with BSE and NSE

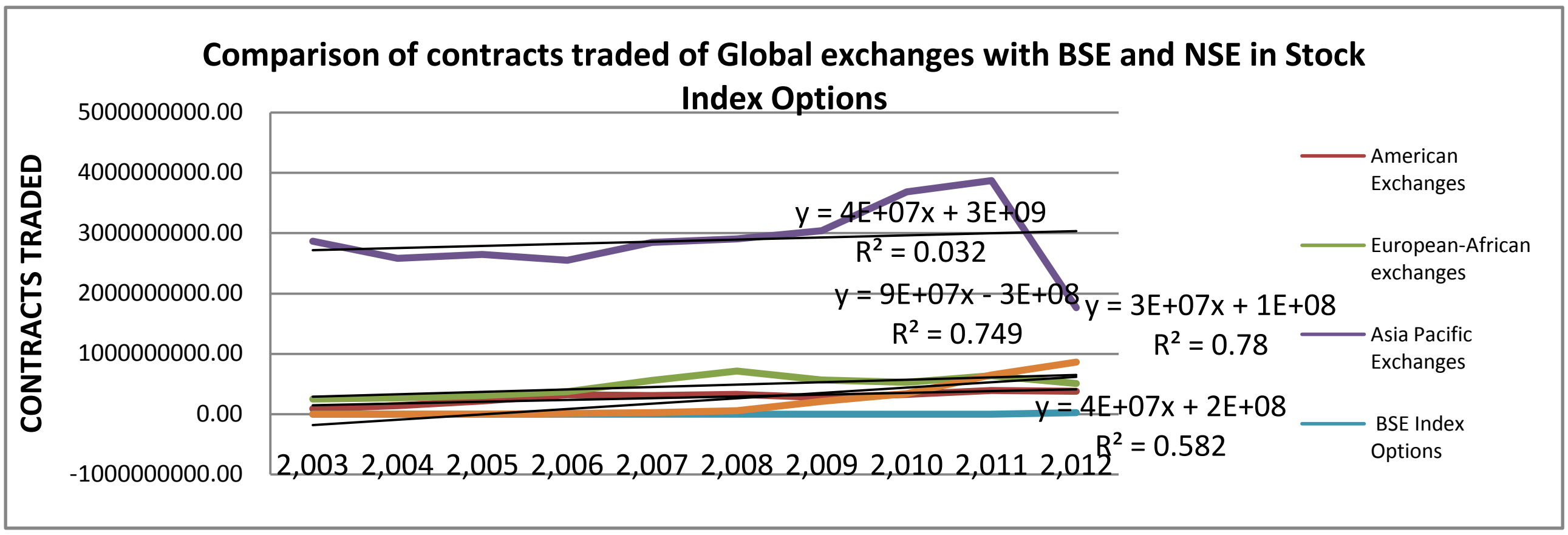

Source: World Federation of Exchanges Graph No: 9

Graph showing the comparison of Stock Options of the Global exchanges with BSE and NSE

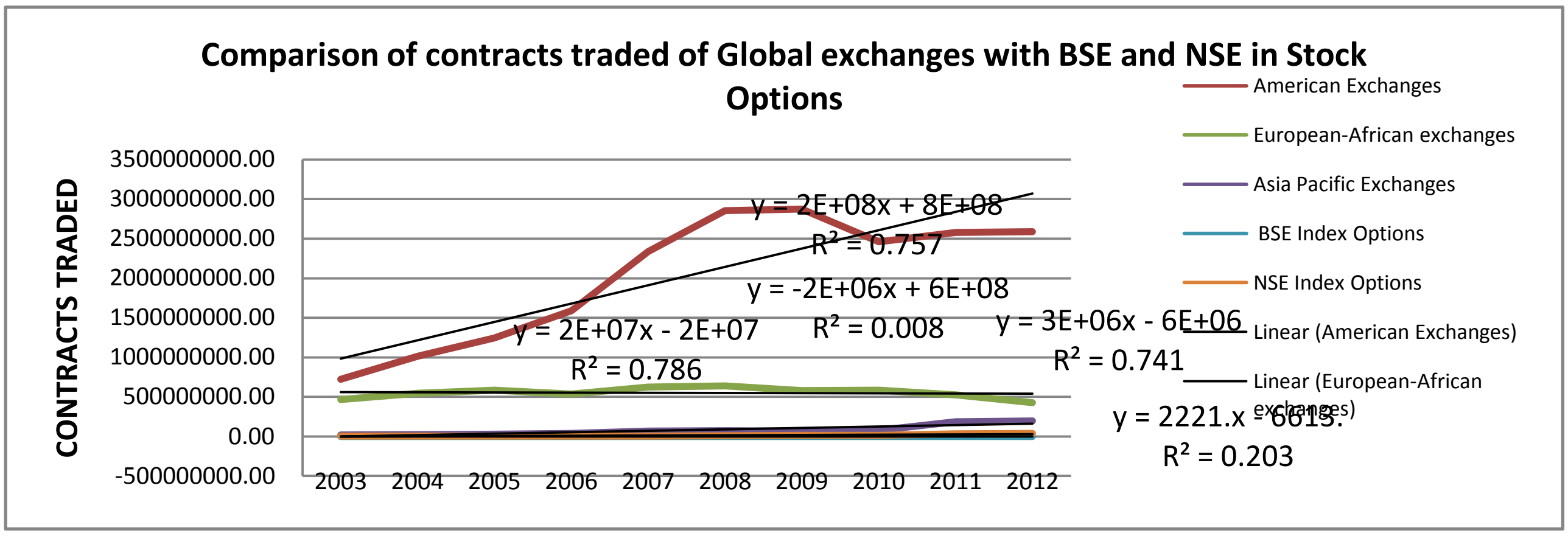

Source: World Federation of Exchanges Graph No: 10

Graph showing the comparison of Stock Futures of the Global exchanges with BSE and NSE

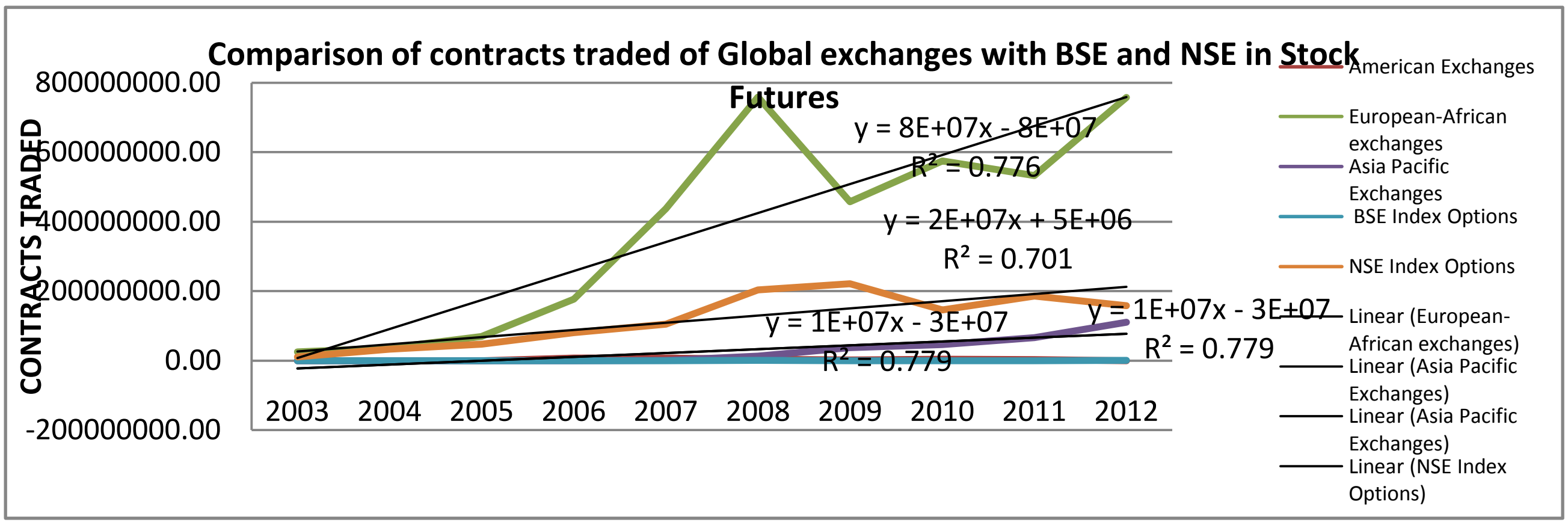

Source: World Federation of Exchanges Graph No: 11

The above four graphs in relation to the comparison of BSE and NSE with Global exchanges suggests that BSE and NSE along with Global exchanges is showing upward trend in respect to all the derivative instruments which is evident from the Regression equation but still way behind in comparison with the growth rate. NSE is still in the comparison mode to the Global exchanges but BSE needs to upscale itself to be on par with the growth rate of the derivative market of the Global exchanges.

**Summary of findings**

1) Derivative market is growing very fast in Indian Economy. The turnover of derivative market is increasing year by year in the India's largest stock exchange NSE.The turnover of NSE is 97 percent higher than BSE. The number of derivative contracts traded in NSE is 75% higher than BSE. NSE meanwhile is ranked third among the top thirty derivative exchanges in terms of number of contracts traded or cleared in the calendar year 2012. Nifty Options have retained their rank as the world's second most traded option in calendar year 2012 as well. And second position respectively. Though the NSE derivative market is much larger compared to the BSE but it lags way behind the global exchanges which rejects the hypothesis that Indian derivatives market is in line with the global exchanges.

2) NSE Derivatives segment surpasses the Cash and Derivative market turnover of not only BSE but also the cash market turnover of its own in daily and yearly trend. There is an impact of introduction of derivatives on Cash market segment as the investors preferring derivatives to hedge their risks in the highly fluctuating price trend in India.

3) For the first time the turnover in the derivatives market of BSE has crossed the equity market turnover in 2011-12 driven mainly by the incentives offered by the exchange. The NSE derivatives turnover is 91 percent higher than equity turnover of its own.

4) Derivatives market segment contribution to GDP ratio is 92 percent higher than the cash market turnover and thus empowering the growth of economy of India.

5) The growth of Index and Stock Futures of BSE in terms of turnover and number of contracts traded is variants with the growth number of contracts and turnover of Index and Stock Futures of NSE and rejecting the hypothesis that there is a correlation within the Index and Stock Futures of BSE and NSE.

6) The Index Call and Put Option of BSE is statistically insignificant which means that average increase in the Index Call and Put Option of NSE may lead to the average increase in the Call and Put Option of NSE in the number of contracts traded but it is statistically significant for the Single Stock Call and Put Option where increase in one does not affect the latter. This partially accepts the hypothesis of the increase in the number of contracts traded for Index Options of BSE leads to the increase in the Index Options of NSE but not the same with the Single Stock Options.

7) The average increase in the turnover of Index Call and Put Option of NSE leads to the increase in the turnover of Index Call and Put Option of BSE which is supported by the p-value of two tailed paired t-test but Stock Call Stock Option of BSE is statistically insignificant with Stock Call Option of NSE supporting the hypothesis that there is a correlation between them but Stock Put Option of BSE is statistically signification with Stock Put Option of NSE with 0.005 value rejecting the hypothesis.

8) Volatility of all the Indices are in line with each other except Nifty Junior which consists of only 50 stock and highly volatile in the beginning compared to other Indices. The volatility has shown a decline in 2000-01 and one of the factor may be due to the introduction of derivatives in India and increase in 2008-09 attributing to the Global crisis supporting the hypothesis that derivatives is one of the factor impacting the volatility of Market.

9) The growth in the number of contracts of Stock Index Option for Asia Pacific and European exchanges have increased considerably but still is variant with the number of contracts traded in American exchanges. In case of Index Future there is a phenomenal increase in the number of contracts traded in Asia Pacific and European exchanges but the American exchanges still showed not a perfect secular trend over the years? European exchanges have surpassed the American exchanges in the year 2012 in the number of Index Future contracts traded.

10) National Stock exchange stands on par with Global exchanges in terms of the number of contracts traded for Stock Index Options and Futures and also Stock Futures but the Stock Options has not made it to the Top 5 exchanges though the options available for trading were changed from American to European style in 2011.

11) Index Options holds the majority in the volume of Contracts traded and turnover in BSE but it differs in the turnover of NSE. The majority of the turnover for NSE is from Stock Options segment with minimal contracts which indicates the high amount traded on each contract on an average over the years and again Index options holds majority in the number of contracts traded even in NSE.

12) BSE Sensex and Derivatives turnover has shown a decline post 2008 could be attributing to the factor of global financial crisis.

13) NSE tops BSE in terms of monthly turnover and number of contracts traded in a relatively high margin in all the derivative instruments rejecting the hypothesis that there is a relation in the increase of BSE and NSE derivative turnover and Number of contracts traded.

14) National stock exchange has made it to the Top 5 exchanges in terms of the contracts traded but when compared with the Total number of contracts traded of all the American exchanges, Asia Pacific and European exchanges, NSE still stands behind but it is ahead of BSE. Hence both BSE and NSE still needs to grow to reach it to the global level which in fact rejects my hypothesis that NSE and BSE are in par with global exchanges in derivatives market.

## Conclusions

In terms of the growth of derivatives markets, and the variety of derivatives users, the Indian market has equaled or exceeded many other regional markets. While the growth is being spearheaded by retail investors, private sector institutions and large corporations, smaller companies and state-owned institutions are gradually getting into the act. Foreign brokers are boosting their presence in India in reaction to the growth in derivatives. The variety of derivatives instruments available for trading is also expanding. In the past, there were major areas of concern for Indian derivatives users. Large gaps exist in the range of derivatives products that are traded actively. In equity derivatives, NSE figures showed that almost 90% of activity was due to index options, index futures & stock futures, whereas trading in options is limited to a few stocks, partly because stock options were of American style & they are settled in cash and not the underlying stocks. But with the start of 2011 all stock options available for trading were changed to European style. This change has led to the liquidity in stock options not only close to ATM strikes but also across multiple strikes just as in case of index options. This change has encouraged the options writers to go ahead eliminating the assignment risk prior to expiry which will eventually benefit them. The study concluded that there is a huge difference in the way BSE and NSE functions in terms of derivatives which is evident with the turnover and the number of contracts traded. The Indian derivatives market has shown a tremendous growth but is still not in line with the global derivatives market. Considering many changes currently, derivatives market in India is poised to grow and mature further to accommodate larger participation across varied asset classes by wide range of participants.

There is a lot of scope of growth for Indian Derivatives Market and it is showing in its signs on the global platform. With the robust regulations, strengthening of our financial structure and more knowledge on derivatives market would gain more of investor confidence in the market and more of trading on derivatives for hedging purpose which is the very purpose for the reason for the existence of derivatives market.

There is a scope to further study the Single Options growth, Currency derivatives, Interest rate derivative, and Credit derivatives in comparison to the global market as they are all in the gowing stages in the Indian derivatives market.